\documentclass[fleqn,10pt]{wlscirep}
\usepackage[utf8]{inputenc}
\usepackage[T1]{fontenc}
\usepackage{lineno}
\usepackage{amsmath}
\usepackage{xcolor}
\usepackage{hyperref}

\newcommand{\irissq}{$IRIS^{2}$}

\title{Chromospheric and Coronal heating in active region plage by dissipation of currents from braiding}
\usepackage{caption}
\usepackage{subcaption}
\usepackage{multibib}
\newcites{Methods}{Additional References}

\DeclareRobustCommand{\ion}[2]{\textup{#1\,\textsc{\lowercase{#2}}}}

\def\Mgk{\hbox{\ion{Mg}{ii}~k}}
\def\kcore{\hbox{\ion{Mg}{ii}~k$_{3}$}}
\def\Si{\hbox{\ion{Si}{iv}}}
\def\HiC{\hbox{\ion{Hi-C}{172}}~\AA}

\author[1,2,3,4,*]{Souvik Bose}
\author[1,3,4]{Bart De Pontieu}
\author[1,2,3,4]{Viggo Hansteen}
\author[1,2]{Alberto Sainz Dalda}
\author[5]{Sabrina Savage}
\author[5]{Amy Winebarger}

\affil[1]{Lockheed Martin Solar \& Astrophysics Laboratory, Palo Alto, CA 94304, USA}
\affil[2]{Bay Area Environmental Research Institute, NASA Research Park, Moffett Field, CA 94035, USA}
\affil[3]{Institute of Theoretical Astrophysics, University of Oslo, PO Box 1029, Blindern 0315, Oslo, Norway}
\affil[4]{Rosseland Centre for Solar Physics, University of Oslo, PO Box 1029, Blindern 0315, Oslo, Norway}
\affil[5]{NASA Marshall Space Flight Center, Huntsville, AL 35812, USA
}
\affil[*]{bose@lmsal.com}

\begin{abstract}

\textbf{It remains unclear which physical processes are responsible for the dramatic increase with height of the temperature in stellar atmospheres, known as the chromospheric ($\sim$10,000 K) and coronal (several million K) heating problems. Statistical studies of sun-like stars reveal that chromospheric and coronal emissions are correlated on a global scale, constraining, in principle, theoretical models of potential heating mechanisms. However, so far, spatially resolved observations of the Sun have surprisingly failed to show a similar correlation on small spatial scales, leaving models poorly constrained.  Here we use unique coordinated high-resolution observations of the chromosphere (from the Interface Region Imaging Spectrograph or IRIS satellite) and low corona (from the Hi-C 2.1 sounding rocket), and machine-learning based inversion techniques to show a strong correlation on spatial scales of a few hundred km between heating in the chromosphere and low corona for regions with strong magnetic field (``plage"). These results are compatible with recent advanced 3D radiative magnetohydrodynamic simulations in which dissipation of current sheets formed due to the braiding of the magnetic field lines deep in the atmosphere is responsible for heating the plasma simultaneously to chromospheric and coronal temperatures. Our results provide deep insight into the nature of the heating mechanism in solar active regions.}
\end{abstract}
\begin{document}
\flushbottom
\maketitle
%
%
\thispagestyle{empty}


\section*{Main}
\label{Section:intro}
It has long been known from statistical studies of a wide variety of stars that the chromospheric and coronal emission are well correlated on a global, stellar scale\cite{1987A&A...172..111S}, which suggests that the heating mechanisms in the chromosphere and corona are related. This intriguing correlation potentially provides critical constraints on the heating mechanism responsible for non-thermal energy dissipation in a stellar atmosphere. This is because the chromosphere and corona are very different in terms of partial ionization, non-local thermodynamic equilibrium (non-LTE), stratification, and magnetic field to plasma pressure ratio. These differences have varying effects on the efficiency and likelihood of different heating mechanisms. 

The global correlations, however, have not been translated into a direct physical connection at high spatial and temporal scales in the solar atmosphere\cite{1999SoPh..190..419D,2003ApJ...590..502D,2015ApJ...809L..30C}, leaving potential heating mechanisms poorly constrained. Detailed correlation studies on small spatial scales between coronal and chromospheric heating are very challenging, because thermal conduction in the corona efficiently redistributes heat along a wide range of heights and locations. De Pontieu et al.\cite{1999SoPh..190..419D,2003ApJ...590..502D} studied the correlation between heating sites in ``moss"-- dynamic, bright regions of enhanced emission observed in the upper transition region (TR) at the footpoints of high-pressure and high-temperature loops above an active region (AR) plage\cite{1999SoPh..190..409B}, and the underlying chromospheric emission (visible in Ca~H~3968~\AA\ filter images from ground-based telescopes). They found no correlation on arcsecond ($\sim$ 700 km) scales, in puzzling contrast with the global, stellar studies. These earlier studies were flawed, however, because recent results indicate that the bright ``chromospheric” emission in broadband Ca~H~3968~\AA\ filter images is not a likely signature of chromospheric heating, but rather emanates from deeper, hotter layers in the convection zone below evacuated strong photospheric magnetic flux tubes\cite{2004ApJ...607L..59K,2006ApJ...646.1405D}. 
Upper TR moss provides an excellent laboratory to study how chromospheric and coronal heating mechanisms are spatio-temporally correlated. Because moss is formed in a thin layer it is very sensitive to changes in the local heating rate and avoids the confusion introduced by line-of-sight superposition that affects optically thin coronal diagnostics\cite{2003ApJ...590..502D,2001ApJ...563..374V}. Most current observations lack the spatial resolution to resolve moss, which is structured on spatial scales of a few hundred km. Recent 
observations from High-Resolution Coronal Imager (Hi-C) flight~1\cite{2014SoPh..289.4393K}, and later Interface Region Imaging Spectrograph (IRIS\cite{2014SoPh..289.2733D}) satellite, showed evidence of highly-localized, small-scale, heating events in active region moss with durations between 15 and 50s\cite{2013ApJ...770L...1T,2020ApJ...889..124T}. This relatively rare short-term, rapid variability is thought to be caused by the impact of non-thermal electrons, generated in the corona by reconnection in nanoflares\cite{1988ApJ...330..474P}, on the upper transition region and lower-lying chromosphere. While these observations hint at a relationship between chromospheric and coronal heating on small scales, these rare events do not seem to be common enough to explain the relatively steady emission patterns in the low atmosphere.

In this paper we use unique coordinated observations of active region moss at unprecedented high spatial and temporal resolution, from IRIS and Hi-C sounding rocket flight 2.1\cite{2019SoPh..294..174R} to investigate the relationship between chromospheric and coronal heating on small spatial scales. We exploit recent advances in modeling of non-LTE radiative transfer of spectral lines formed in the chromosphere (refs.\cite{2013ApJ...772...90L,2015ApJ...809L..30C}) that indicate that the cores of the chromospheric \ion{Mg}{ii}~h\&k lines in moss are well-defined signatures of chromospheric heating, in contrast to broadband \ion{Ca}{ii}~H filter images. The combination of state-of-the-art spectral inversions of such lines, combined with the high spatio-temporal resolution of the 5-min timeseries of Hi-C~2.1 images allow us to determine unequivocally whether chromospheric and coronal heating are spatially and temporally correlated on small spatial scales, thereby providing strict constraints for heating based models. 

\begin{figure*}[ht!]
\centering
\includegraphics[width=\linewidth,trim=0cm 12cm 0.cm 3.cm,clip,height=13.5cm]{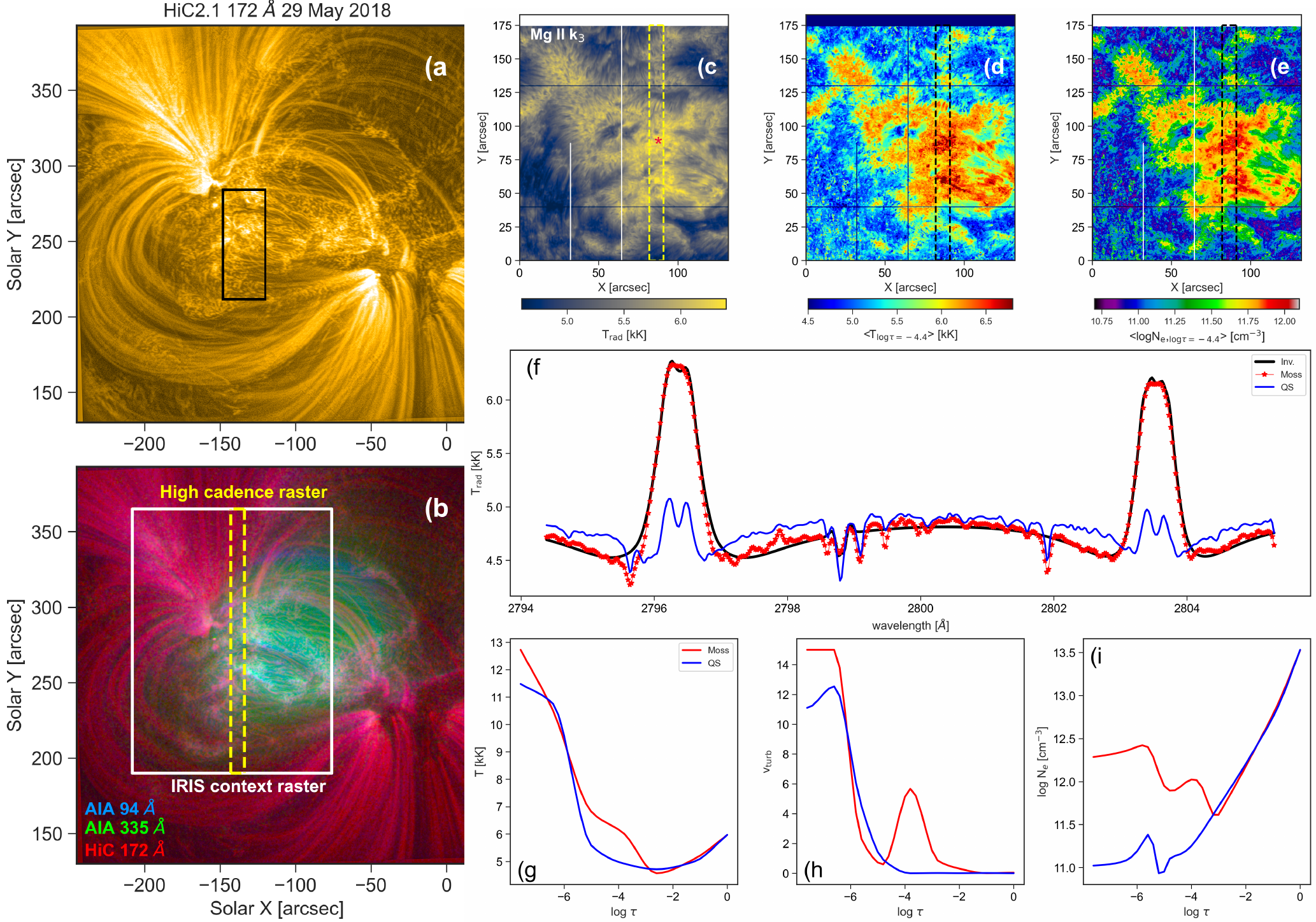}
\caption{\textbf{The chromospheric and coronal scenery of the moss observed with HiC-2.1 and IRIS on 29 May 2018.} (a)~\HiC{} observation of AR 12712 centered at solar ($X$,$Y$)=(-145",280"). The black rectangular box outlines the moss region studied in this paper. (b)~A composite layering showing \ion{Fe}{iX/x} \HiC{} (red, 1~MK), \ion{Fe}{xvi} 335~\AA\ (green, 2.5~MK) and \ion{Fe}{xviii} 94~\AA\ (blue, 6~MK) images from AIA. The white and yellow-dashed boxes indicate the FOV recorded in the IRIS context and high-cadence rasters respectively. (c)~\kcore{} emission (T$_{\mathrm{rad}}$) from the context raster. (d) and (e) chromospheric plasma temperature ($T$) and electron density (N$_{\mathrm{e}}$) maps corresponding to the context raster averaged between log$\tau$=[$-$4.6, $-$4.2]. The FOV of the high-cadence raster is outlined in the three panels. (f) Observed (red) and inverted (black) \ion{Mg}{ii}~h\&k profiles corresponding to a moss pixel indicated in panel~(c). A typical quiet-Sun profile is shown (blue) for reference. Panels~(g), (h), and (i) show the stratification of $T$, microturbluent velocity, and N$_{\mathrm{e}}$ of the moss (red) and quiet-Sun (blue) inferred from \irissq\ inversion. }
\label{fig:context}
\end{figure*}

Figure~\ref{fig:context} shows the coronal and the chromospheric emission of the AR target 12712. The moss region, identified as a reticulated bright emission patch in panel~(a), is clearly seen to lie at the footpoints of the hotter ($\sim$2--6~MK, see panel~b and Materials and Methods) AR loops, typical of a moss\cite{1999SoPh..190..409B,1999SoPh..190..419D,2003ApJ...590..502D,2010A&A...518A..42T}.
IRIS co-observed the AR target, and we utilize the spectral information primarily in the \ion{Mg}{ii}~h\&k lines to explore the chromosphere underneath the moss which is characterized by regions of enhanced brightness in the \kcore{} (h$_{3}$) core. In panel~(c), such a region is indicated by the area bounded between $X$~$\sim$~80--100" and $Y$~$\sim$~40--100", and it corresponds to a chromospheric plage. This observation is in agreement with Carlsson et al.\cite{2015ApJ...809L..30C} who also reported that the \Mgk{} spectral line in plages show little-to-no reversals of the upper chromospheric k$_{3}$ core. Based on a semi-empirical radiative transfer approach, they argued that the enhanced core brightness in the plage represents conditions of hot and dense chromosphere, which causes the source function to remain strongly coupled with the local Planck function, thereby indicating locally strong heating.

Instead of an ad-hoc forward modeling approach, for the first time, we employ a state-of-the-art, simultaneous, multi-line inversion scheme based on \irissq\cite{2019ApJ...875L..18S} with six different spectral lines (including \ion{C}{ii}~1335~\AA, refer: Materials and Methods). Figures~\ref{fig:context}~(d) and (e) show the plasma temperature ($T$) and electron density (N$_{\mathrm{e}}$) maps derived from the inversion scheme, which upon inspection, immediately reveals clear evidence of hot and dense plasma in the regions associated with enhanced \kcore{} radiation temperature (T$_{\mathrm{rad}}$, panel~c). The radiation temperature is the temperature for which the Planck function reproduces the observed intensity at any given wavelength. Interestingly, the moss region (identified from panels a and b) show stronger enhancement in the \kcore{} radiation temperature that appears to be well correlated with stronger $T$ and N$_{e}$ values, which clearly suggests that the chromosphere underneath the moss has hotter and denser plasma associated with it. Including multiple atomic species in the inversion allows us to effectively constrain a wide range of solar atmosphere spanning the photosphere through the upper chromosphere (see Sainz Dalda et al.~2022, in prep.).
This approach represents a major advancement since it enables to directly examine and correlate the physical parameters of the plasma, instead of focusing on the ``proxies" of heating (as had been done in the past\cite{2001ApJ...563..374V,2003ApJ...590..502D}).

\begin{figure}[ht!]
\centering
\includegraphics[scale=0.5]{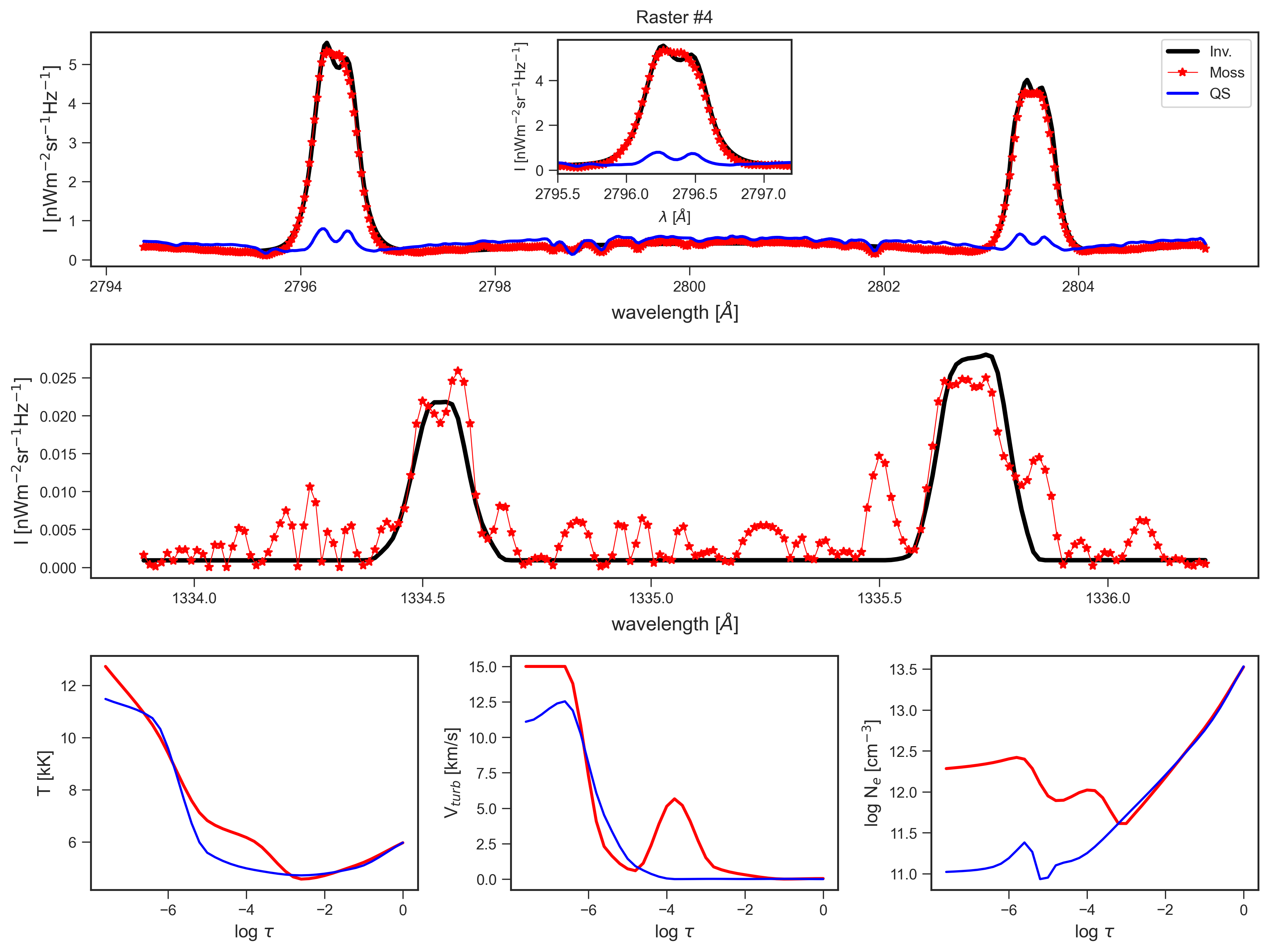}
\caption{Comparison of the observed and fitted \ion{Mg}{ii}~h\&k (top row) and \ion{C}{ii}~1335~\AA\ (middle row) spectra for a mossy plage pixel from IRIS raster \#4. The derived stratification of T, microturbulent velocity (V$_{\mathrm{turb}}$), and N$_{\mathrm{e}}$ and their comparison with a typical QS atmosphere is shown in the bottom row.}
\label{fig:supp_inv_ex1}
\end{figure}

\begin{figure}[ht!]
\centering
\includegraphics[scale=0.5]{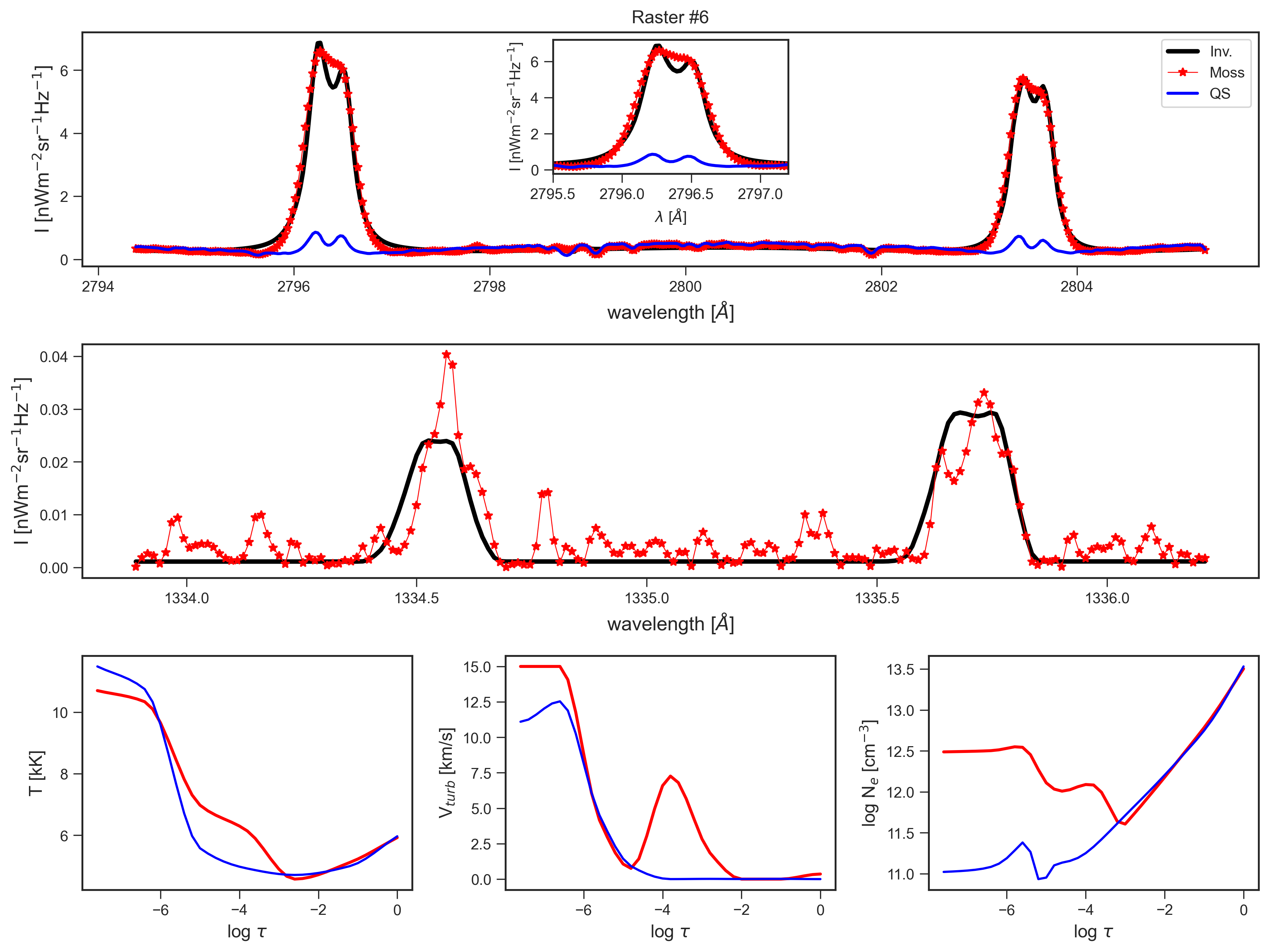}
\caption{Second example from IRIS raster \#6 showing the fits between the observed and the synthetic spectra for a pixel in a moss region, in the same format as Supp. Fig.~\ref{fig:supp_inv_ex1}.   }
\label{fig:supp_inv_ex2}
\end{figure}

\begin{figure}[ht!]
\centering
\includegraphics[scale=0.5]{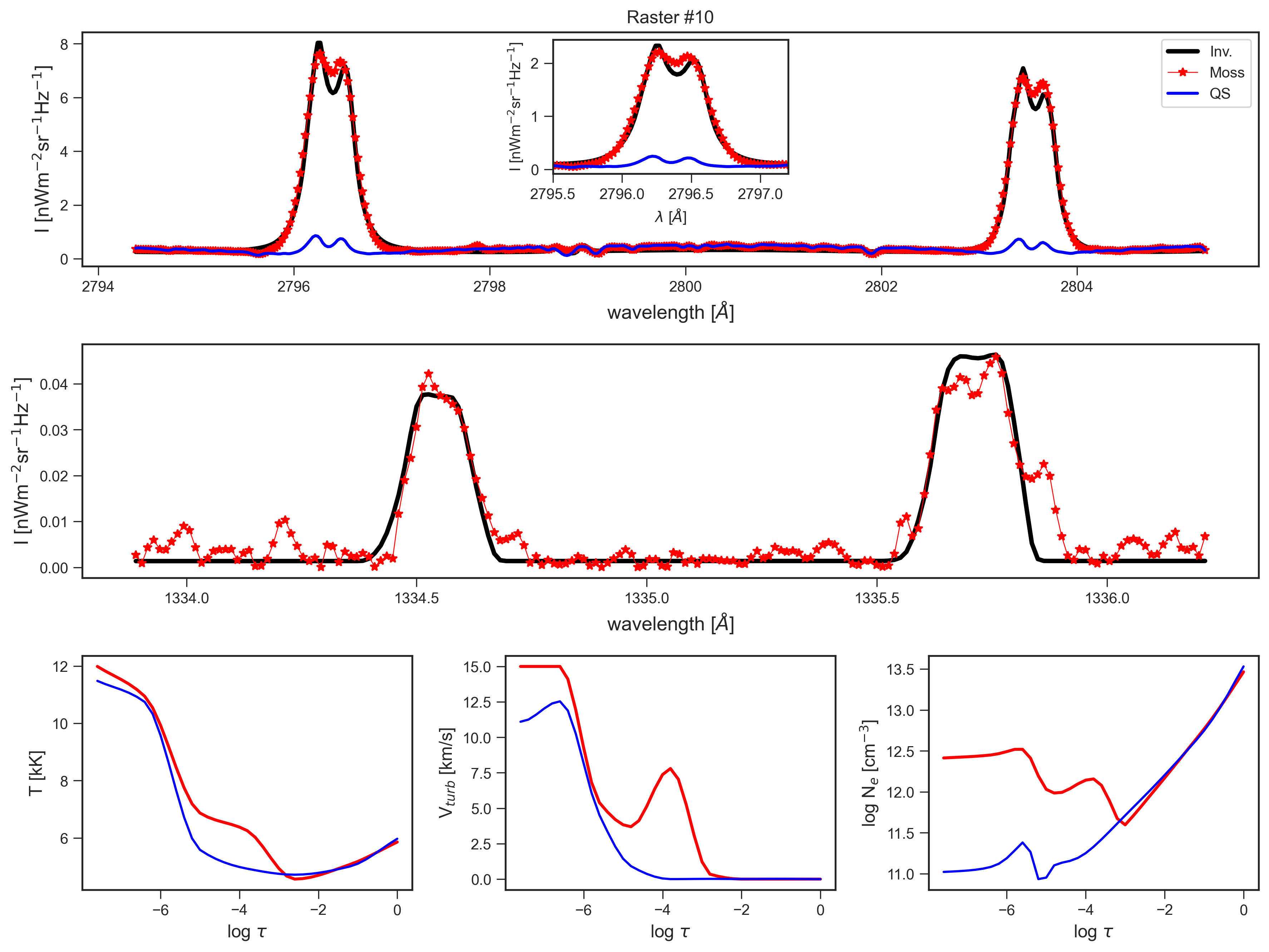}
\caption{Third example from IRIS raster \#10 showing the fits between the observed and the synthetic spectra for a pixel in a moss region in the same format as Supp. Fig.~\ref{fig:supp_inv_ex1}. }
\label{fig:supp_inv_ex3}
\end{figure}

A representative example of a \ion{Mg}{ii}~h\&k spectrum associated with a mossy plage pixel is shown in Fig.~\ref{fig:context}~(f). The spectrum shows a clear lack of self-reversal in the k$_{3}$ (h$_{3}$) wavelength intensities indicating a flat-topped (single-peaked\cite{2015ApJ...809L..30C}) profile unlike the quiet-Sun (QS) profiles, which show a distinct absorption in k$_{3}$ (h$_{3}$). Moreover, the overall T$_{\mathrm{rad}}$ and the width of the plage spectrum is significantly enhanced compared to the QS case, which is in agreement with earlier findings\cite{2015ApJ...809L..30C}. The synthetic spectra derived from inversion show good fits not only for the example shown in Fig.~\ref{fig:context}~(f), but also for various other examples shown in Supp.~Figs.~\ref{fig:supp_inv_ex1}, \ref{fig:supp_inv_ex2}, and \ref{fig:supp_inv_ex3}. Additionally, the \ion{C}{ii} lines, which were simultaneously inverted and fitted with the \ion{Mg}{ii}~h\&k, show good fits as well; this suggests that the derived physical parameters are well-constrained at upper chromospheric heights. The inferred $T$, microturbulence and N$_{\mathrm{e}}$ as a function log$\tau$ ($\tau$ being an ``optical" measure of depth in the solar atmosphere) corresponding to the plage and the QS spectra in panel~(f) is shown in panels~(g)--(i) of Fig.~\ref{fig:context}. Interestingly, all the three parameters show a clear enhancement centered around log$\tau$=$-$4 compared to the QS case, reinforcing that the chromosphere underneath the moss is hot and dense, and has significant non-thermal velocities (of $\sim$6~kms$^{-1}$, here named as microturbulence). This pattern is consistently observed for a vast majority of the plage including the examples shown in Supp.~Figs.~\ref{fig:supp_inv_ex1}, \ref{fig:supp_inv_ex2}, and \ref{fig:supp_inv_ex3}. In addition, the temperature stratification shows a characteristic ``step" (around log$\tau$=$-$4.5), and the temperature minimum is pushed deeper in the solar atmosphere, which is critical (in combination with high density) in reproducing the flat-topped, single-peaked spectra. 

\begin{figure}[ht!]
    \centering
    \begin{minipage}{0.49\textwidth}
        \centering
        \includegraphics[width=0.9\textwidth,height=16cm]{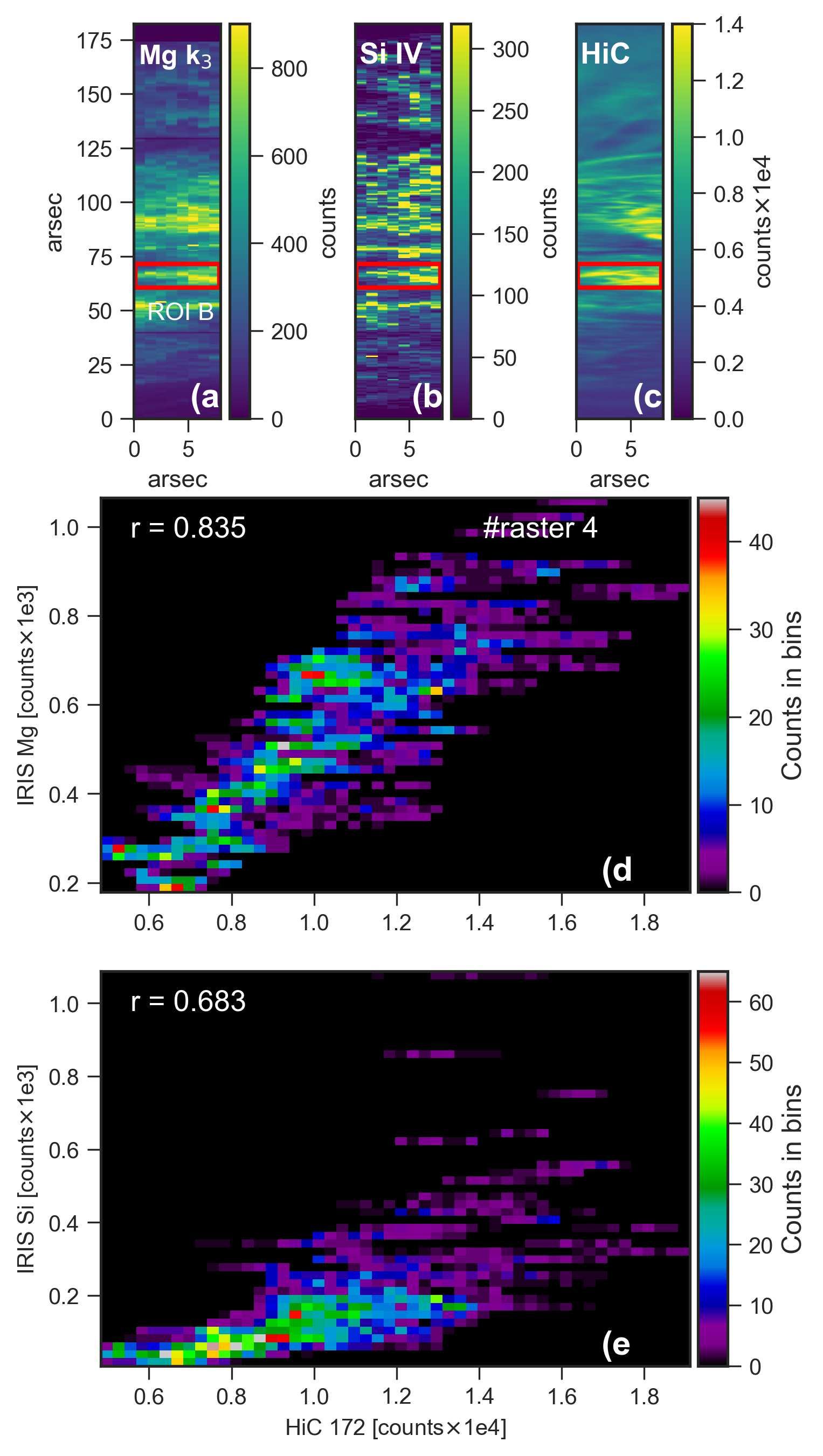} 
    \end{minipage}\hfill
    \begin{minipage}{0.49\textwidth}
        \centering
        \includegraphics[width=0.9\textwidth,height=16cm]{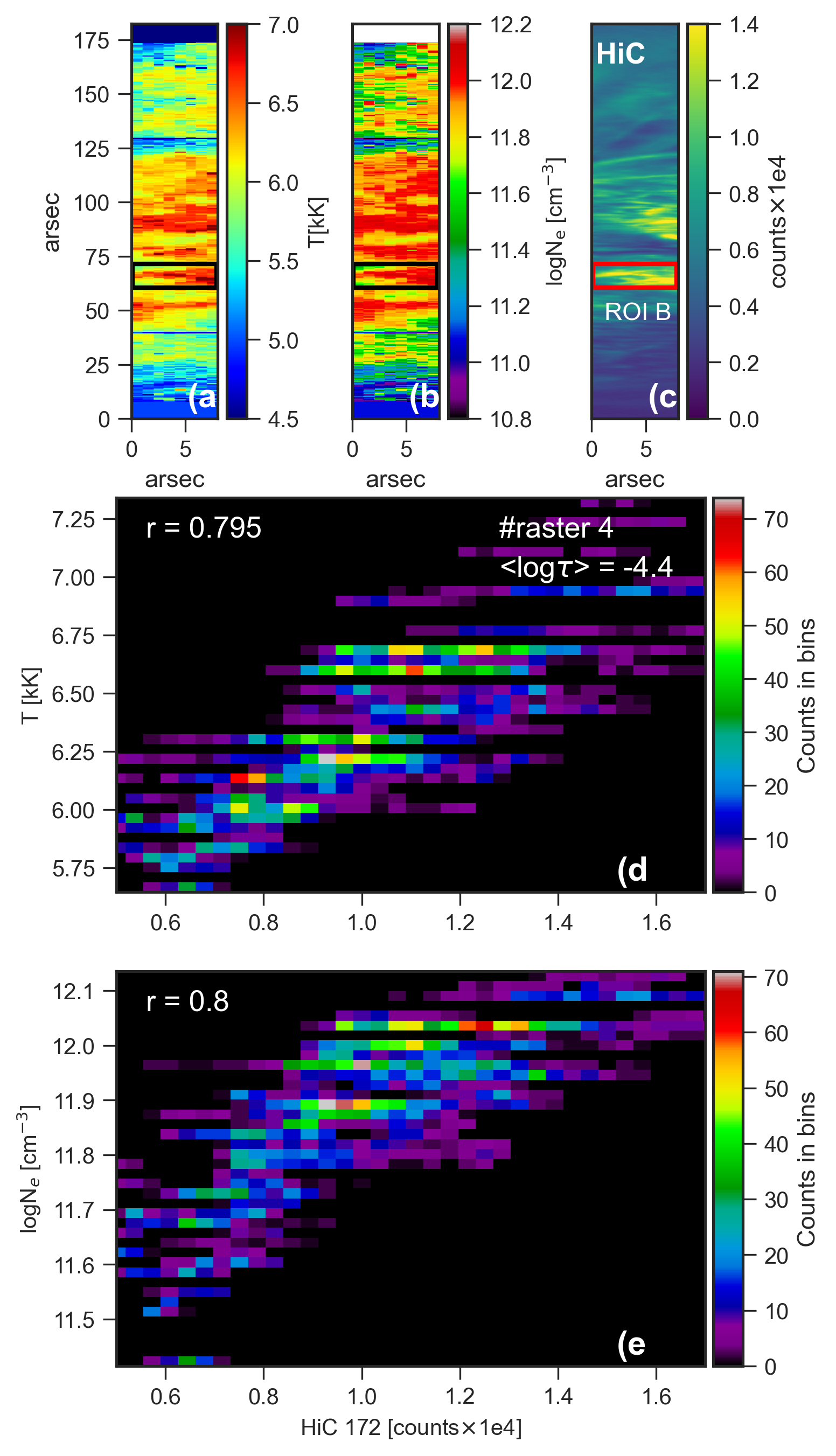} 
    \end{minipage}
    \caption{\textbf{ Spatio-temporal correlation for the mossy plage region ROI-B.} \textit{Left:} (a), (b), and (c) show the \kcore{}, \Si{} and \HiC{} intensity (counts) corresponding to the high-cadence raster (\#4). The FOV under investigation is shown in red. (d) and (e) show the 2D density distribution functions between \kcore{} and \HiC{}, and \Si{} and \HiC{} intensities. \textit{Right:} (a) and (b) show the $T$ and N$_{\mathrm{e}}$ maps corresponding to the same time and FOV as left panel averaged between log$\tau$=[$-$4.6,~$-$4.2], while (c) shows the \HiC{} as a reference. (d) and (e) show the 2D density distribution functions between $T$ and \HiC{}, and N$_{\mathrm{e}}$ and \HiC{}. The Pearson correlation coefficients ($r$) are shown on top left. An animation of this figure is \href{https://www.dropbox.com/sh/c8bbi6aab6srguj/AABCLg8HVVw6GSFhcBfAuol9a?dl=0}{available}.}
    \label{fig:int_therm_corr-B}
\end{figure}

\begin{figure}[ht!]
    \centering
    \begin{minipage}{0.49\textwidth}
        \centering
        \includegraphics[width=0.9\textwidth,height=16cm]{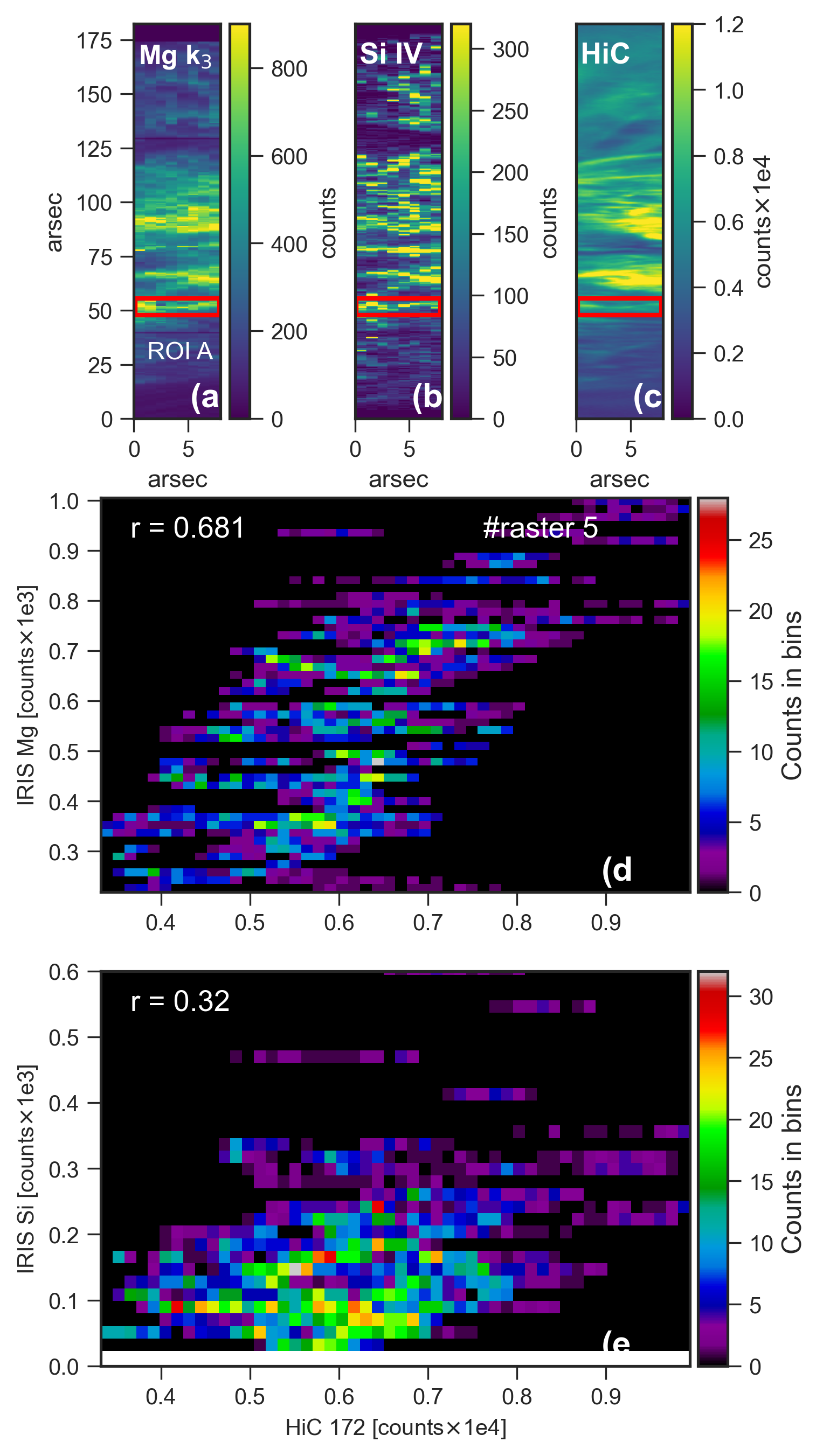} 
    \end{minipage}\hfill
    \begin{minipage}{0.49\textwidth}
        \centering
        \includegraphics[width=0.9\textwidth,height=16cm]{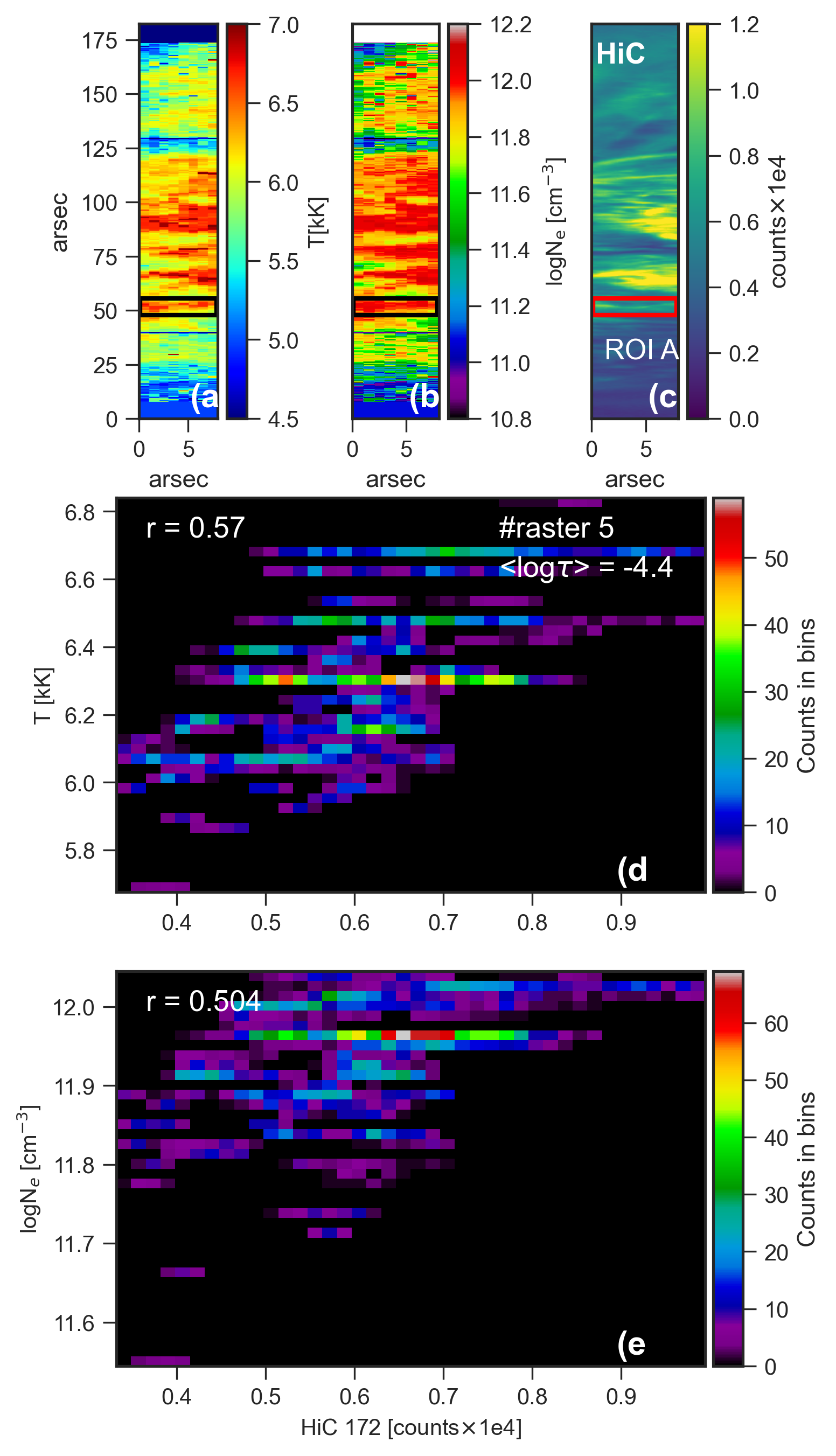} 
    \end{minipage}
    \caption{Same as Fig.~\ref{fig:int_therm_corr-B} but for ROI-A.}
    \label{fig:int_therm_corr-A}
\end{figure}

\begin{figure}[ht!]
    \centering
    \begin{minipage}{0.49\textwidth}
        \centering
        \includegraphics[width=0.9\textwidth,height=16cm]{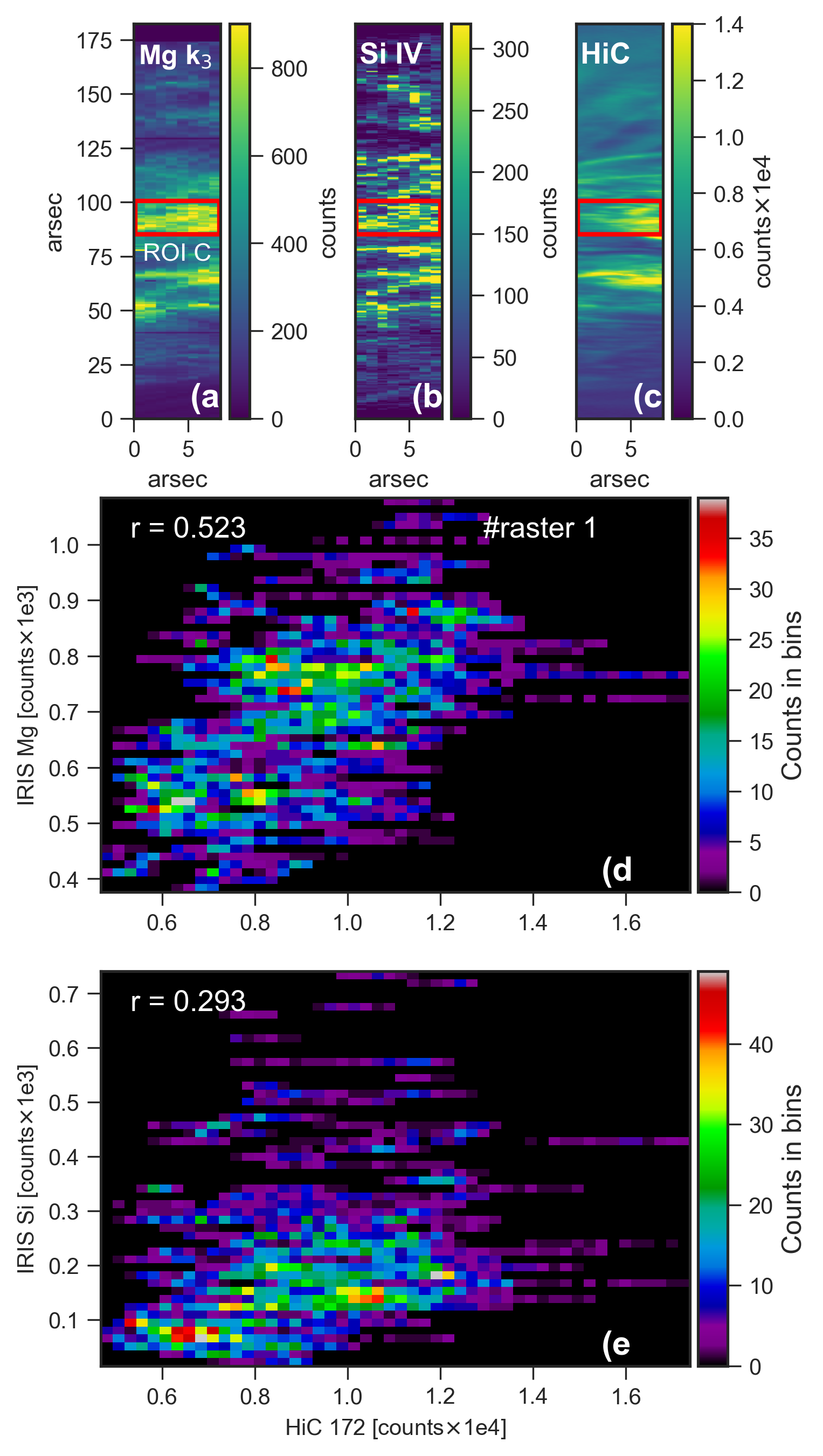} 
    \end{minipage}\hfill
    \begin{minipage}{0.49\textwidth}
        \centering
        \includegraphics[width=0.9\textwidth,height=16cm]{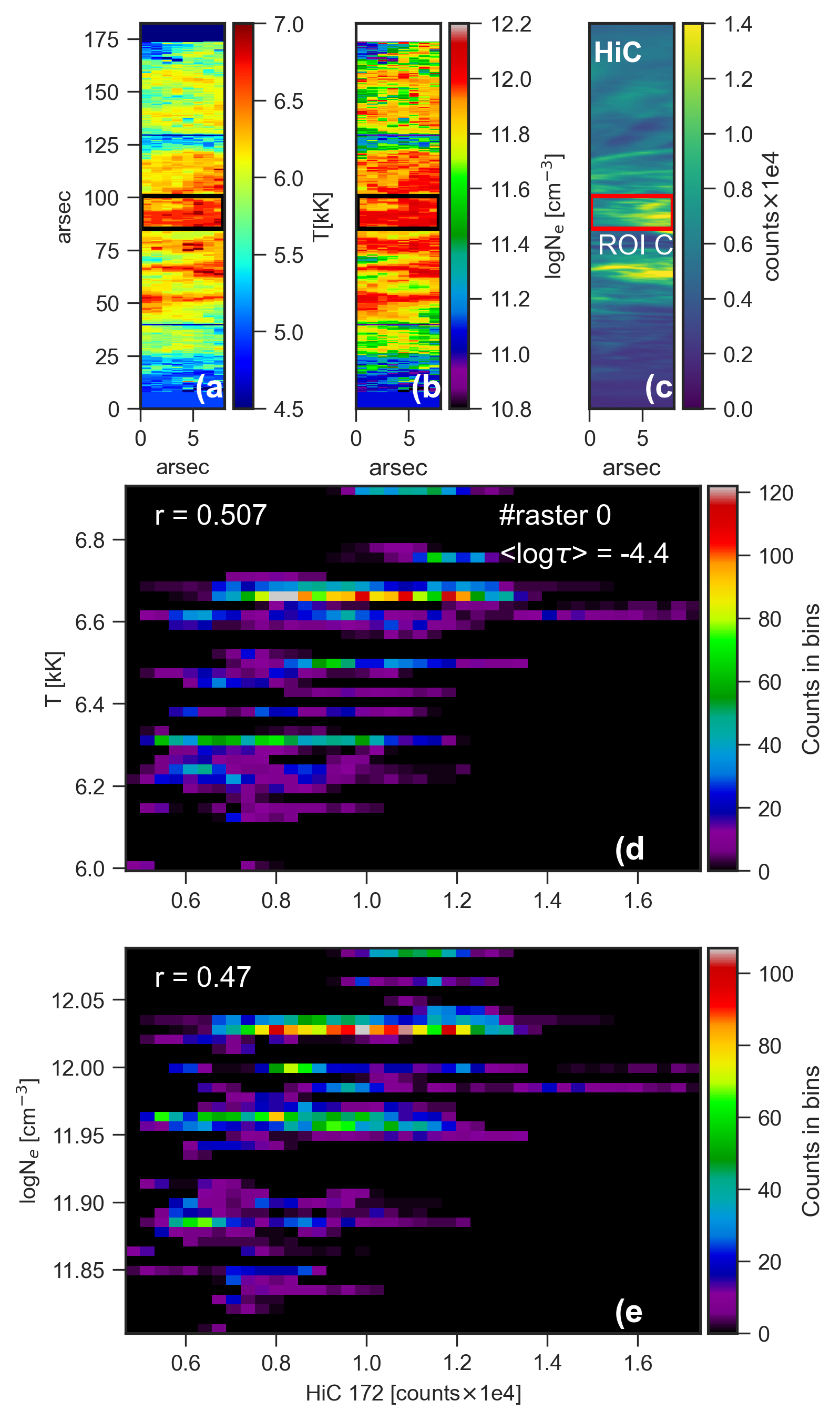} 
    \end{minipage}
    \caption{Same as Fig.~\ref{fig:int_therm_corr-B} but for ROI-C.}
    \label{fig:int_therm_corr-C}
\end{figure}

The left and the right panels of Fig.~\ref{fig:int_therm_corr-B} show the spatio-temporal correlation analysis between the IRIS and the Hi-C intensities, and between the physical parameters derived from the \irissq\ inversion and Hi-C intensity for a small field-of-view (FOV, of 13$\times$7~arcsec$^{2}$) of the moss region marked as the region of interest (ROI)-B. The analysis is derived from the high-cadence IRIS rasters (see Materials and Methods) that overlaps the moss region as shown in Fig.~\ref{fig:context}~(b). Visual inspection reveals that the intensity in ROI-B is well correlated among the \kcore{}, \Si{} and Hi-C channels. This is supported quantitatively in the form of 2-dimensional (2D) density distribution functions shown in panels~(d) and (e), where we see a tight correlation not only between \kcore{} and \HiC{} intensities, but also between \Si{} and \HiC{}. The high values of Pearson coefficient ($r$) in both cases is in direct contrast with earlier results (such as refs.\cite{1999SoPh..190..419D,2003ApJ...590..502D}) where the authors failed to obtain a good correlation on (sub)arcsecond scales. The correlation ($r$) between \Si{} and \HiC{} intensity is slightly worse than the chromospheric (\kcore{}) counterpart (also see Supp. Figs.~\ref{fig:int_therm_corr-A} and \ref{fig:int_therm_corr-C}), possibly because TR \Si{} is often impacted by shocks\cite{2015ApJ...803...44M} and spicules\cite{2003ApJ...590..502D,2006ApJ...646.1405D,2021A&A...654A..51B} that affect its brightness variation more significantly compared to \Mgk{}. Nonetheless, the correlation coefficients obtained in this paper are significantly better than past findings which implies a tight coupling in the heating of the solar atmosphere from chromosphere to coronal heights.

The tight coupling in the moss intensity among the different channels directly translated to a strong visual correlation between the inferred physical parameters (Fig.~\ref{fig:int_therm_corr-B}, right panel), i.e. $T$ (panel~a) and N$_{\mathrm{e}}$ (panel~b) shown at <log$\tau$>=$-$4.4,\footnote{<> represents an average of the physical parameter between log$\tau$=[$-$4.6,$-$4.2]} with \HiC{} intensity (panel~c). In ROI-B, the enhanced \HiC{} intensity is clearly characterized by enhanced $T$ and N$_{\mathrm{e}}$. The range of the derived $T$ and N$_{\mathrm{e}}$ is consistent with values that are typically observed in a chromospheric plage\cite{2015ApJ...809L..30C,2020A&A...634A..56D}. The 2D density distribution functions shown in panels~(d) and (e) further highlight the rigid coupling of the chromospheric and (lower) coronal plasma in a moss. Moreover, the animation associated with Fig.~\ref{fig:int_therm_corr-B} suggests that the correlation remains stable (and high) for the entire timeseries of the Hi-C data. This suggests that the heating observed in ROI-B (and also over the whole moss in general) is roughly constant and non-impulsive (unlike electron beams which causes a sudden, temporal variability of $\sim$15--60s\cite{2013ApJ...770L...1T,2020ApJ...889..124T}). Analysis of the spatio-temporal correlation analysis of other ROIs in the moss region (see Supp. Figs.~\ref{fig:int_therm_corr-A} and \ref{fig:int_therm_corr-C}) reveal similar results ( with slightly different values of $r$), providing support for the non-impulsive nature of chromospheric and coronal heating.

\begin{figure}[ht!]
\centering
\includegraphics[scale=0.55]{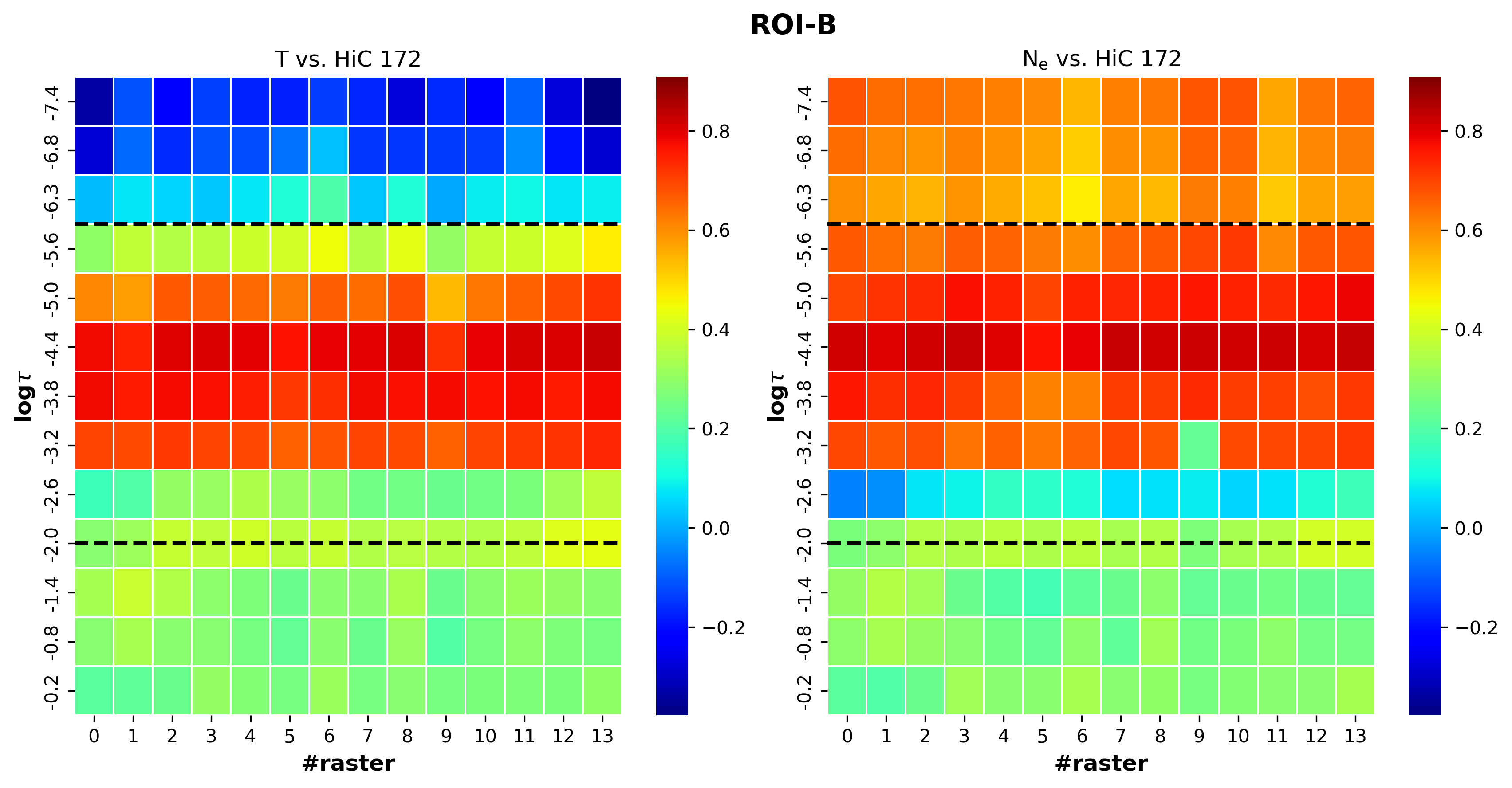}
\caption{\textbf{Heatmaps showing the depth dependent correlation for ROI-B.} \textit{Left:} Pearson correlation coefficient ($r$, encoded in a rainbow color scheme) as a function of depth in the solar atmosphere (log$\tau$, y-axis) and time (\#raster, x-axis) for $T$ vs. \HiC{} and \textit{right:} N$_{\mathrm{e}}$ vs. \HiC{} intensity. The T and N$_{\mathrm{e}}$ are derived from the multi-line inversions of the chromospheric IRIS data and the region between the dashed lines in the two panels indicate the range of optical depths within which the inversions are sensitive (refer to main text). The number of rasters correspond to the entire Hi-C 2.1 duration of $\sim$5~min. }
\label{fig:heatmap-B}
\end{figure}

\begin{figure}[htb!]
\centering
\includegraphics[width=0.81\textwidth,height=18.55cm]{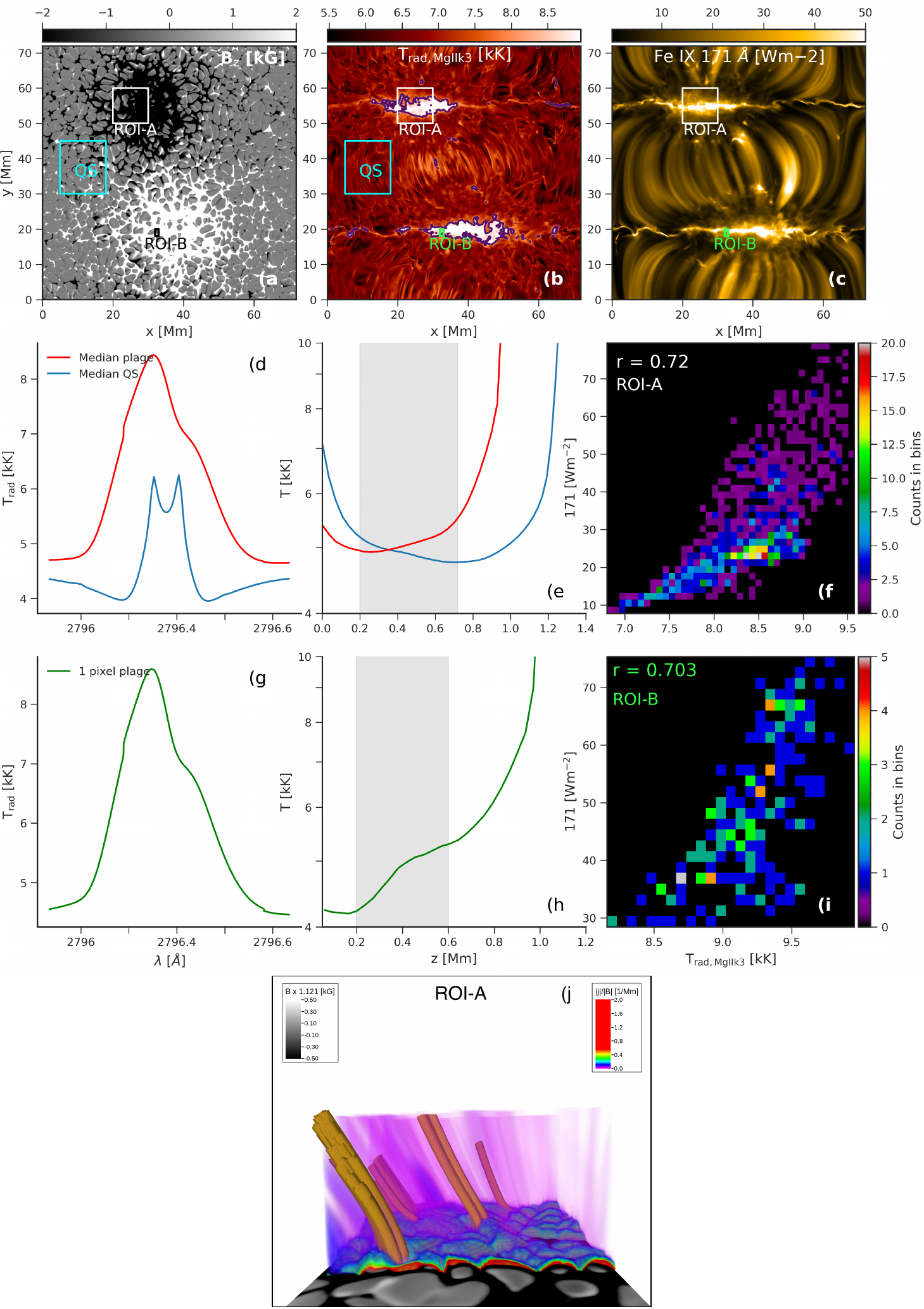}
\caption{\textbf{Overview of the 3D numerical simulation of an active region plage.} (a) Vertical magnetic field strength (B$_{\mathrm{z}}$) at $z$=0~Mm saturated between $\pm$~2~kG. ROIs-A and B cover an area of 10$\times$10~Mm$^{2}$, and 1.5$\times$1.5~Mm$^{2}$ respectively, whereas the quiet-Sun (QS) patch encloses an area of 15$\times$15~Mm$^{2}$. (b) and (c) Synthetic \kcore{} and \ion{Fe}{ix}~171~\AA\ maps of the simulated plage. Purple contours in (b) indicate the regions of enhanced brightness corresponding to the moss emission in (c). (d) Shows the synthetic median \Mgk{} spectra over the mossy pixels (red) within the purple contours in (b) and the QS patch (blue), while (e) shows the corresponding temperature stratification. (g) and (h) shows illustrative examples of a synthetic \Mgk{} spectrum and its ``step-like" temperature stratification (highlighted in gray) for a single mossy plage pixel in ROI-B. (f) and (i) show the 2D density distribution functions for ROIs-A and B. (j) shows the visualization of the magnetic field lines (yellow) and their associated current sheets ($|\Vec{j}|/|\Vec{B}|$, $\Vec{j}$ being the current density) from the photosphere to the corona in ROI-A. }
\label{fig:sim_analysis}
\end{figure}

\begin{figure*}[ht!]
\centering
\includegraphics[scale=0.55]{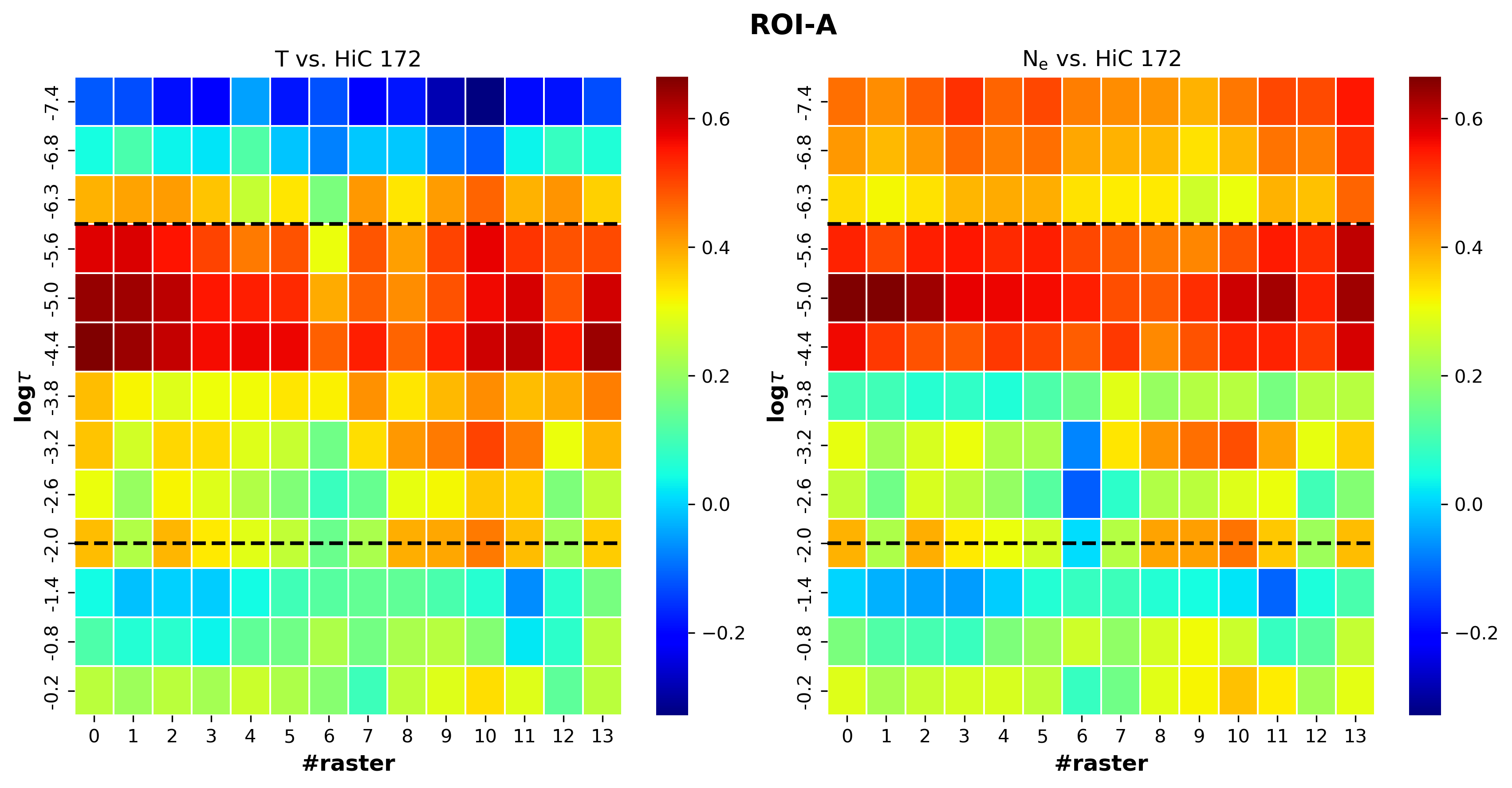}
\caption{Heatmap showing the depth dependent correlation coefficients for ROI-A in the same format as Fig.~\ref{fig:heatmap-B}. }
\label{fig:heatmap-A}
\end{figure*}

\begin{figure*}[ht!]
\centering
\includegraphics[scale=0.55]{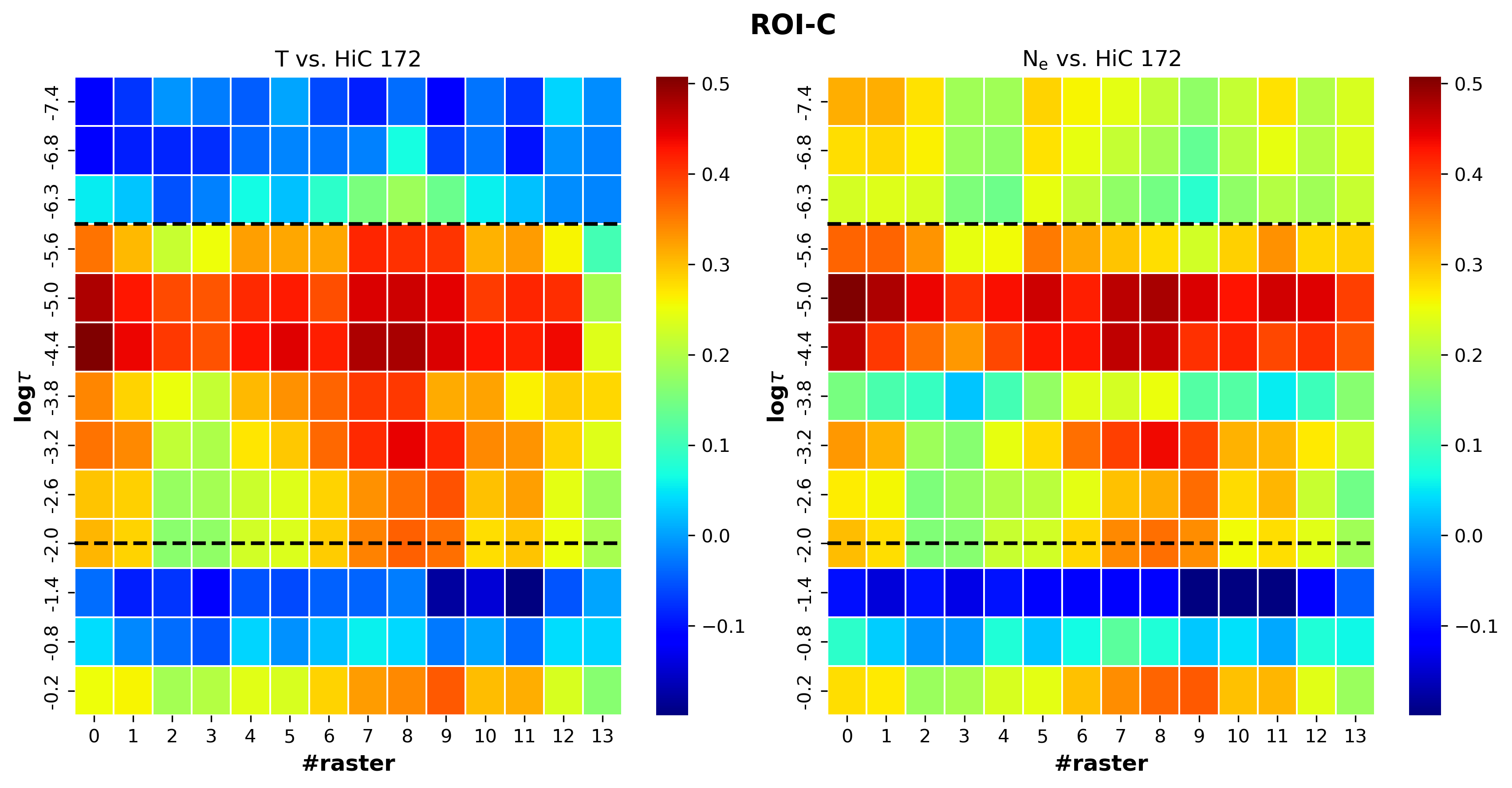}
\caption{Heatmap showing the depth dependent correlation coefficients for ROI-C in the same format as Fig.~\ref{fig:heatmap-B}. }
\label{fig:heatmap-C}
\end{figure*}

Although we found a good correlation between temperature, density (at mid-chromospheric heights), and \HiC{} intensity, an important question is how deep in the solar atmosphere does the correlation still hold? 
The \irissq\ database provides the derived thermodynamic parameters as a function of optical depth $\tau$. The latter is correlated with height, with smaller depths occurring at greater heights. We therefore studied the correlation over a range of heights in the solar atmosphere between log$\tau$=[$-$7.6,0], i.e., from the top of the chromosphere down to the photosphere. This is represented in the form of a 2D heatmap in Fig.~\ref{fig:heatmap-B} that shows the Pearson~$r$ for ROI-B as a function of time (\#raster) and optical depth (log$\tau$) for $T$ (left panel) and N$_{\mathrm{e}}$ (right panel) vs. \HiC{} intensity. The sensitivity of our inversions is predominantly limited between $-$6~$\le$~log$\tau$~$\le$~$-$2 (due to the sensitivities of the different lines included in the inversion, see Materials and Methods and Sainz Dalda et al.~2022~in prep). Interestingly, we find that there is a range of optical depths (within log$\tau$=[$-$6,$-$2]) for which $r$ is rather high ($\ge$~0.6) in both panels (Fig.~\ref{fig:heatmap-B}). The temperature and density of the chromospheric plasma appear to be well correlated with the coronal intensity of \HiC{} down to the temperature minimum (log$\tau$$\sim$$-$3) for all the 14 rasters. The coupling immediately weakens deeper in the atmosphere (for log$\tau$~>~$-$3 in the photosphere) suggesting different physical conditions of the plasma. However, there is no doubt that hot and dense plasma dominates the lower to upper chromosphere underneath, and is well correlated with lower coronal moss brightness. For heights beyond log$\tau$=$-$6, the apparent absence (or presence) of correlation in Fig.~\ref{fig:heatmap-B} does not indicate a lack (or sufficiency) of coupling. It merely implies that the inferred values of T and N$_{\mathrm{e}}$ at these heights are ill constrained due to little or no sensitivity of the inversions. We find similar results for the other ROIs (with different peak values of $r$, Supp. Figs.~\ref{fig:heatmap-A} and \ref{fig:heatmap-C}) where the correlation extends reasonably well down to at least log$\tau$$\sim$$-$4 where the temperature is between 5--6~kK. 

The tight correlation between the derived physical parameters with the \HiC{} intensity down to the temperature minimum strongly suggests the prevalence of a \textit{common} heating mechanism. Thermal conduction, which obviously plays a role in the hot corona and upper transition region,\cite{2001ApJ...563..374V} 
is negligible at low temperatures in the lower chromosphere ($\sim$~6kK), and thus cannot explain the observed spatio-temporal correlation between low chromospheric and upper transition region heating. 
Accelerated electron beams generated in coronal nanoflares can have an impact on heating the chromosphere\cite{2020A&A...643A..27F}, however, due to their impulsive nature such heating would show a high temporal variability\cite{2013ApJ...770L...1T,2020ApJ...889..124T}, which is not observed in this study.

We find that our observations show a striking correspondence with an advanced 3D magnetohydrodynamic (MHD) numerical simulation of active region plage (Fig.~\ref{fig:sim_analysis}~a). The numerical domain encompasses the convection zone, photosphere, chromosphere, TR, and corona (see Materials and Methods). Optically thin modeling of the coronal channels reveal the presence of moss-like emission patches in \ion{Fe}{iX}~171~\AA\ that lie at the footpoints of hot loops visible in \ion{Fe}{xvi}~335~\AA\ (Supp. Fig.~\ref{fig:sim_hot_loops}). Synthetic \kcore{} and 171~\AA\ (panels~b and c) intensity bear a close resemblance with observations where the moss coincides with enhanced k$_{3}$ radiation, and the \Mgk{} spectra show single peaks (panels~d and g). The temperature minimum is shifted deeper in the solar atmosphere for the moss pixels (panel~d) with many showing characteristic ``steps" around low-chromosphere at $z$=0.4~Mm (panel~h), similar to Fig.~\ref{fig:context}~(f). 2D density distribution functions between 171~\AA\ and k$_{3}$ channels show a tight correlation with high values of Pearson $r$ at both large (ROI-A, panel~f) and small (sub)arcsecond scales (ROI-B, panel~i). The simulated scenario bears a close resemblance with observations, and a detailed analysis of the magnetic field topology reveals the presence of braiding in the field lines\cite{1983ApJ...264..635P} deep in the atmosphere, which leads to the formation of current sheets (panel~j). These sheets extend from the chromosphere (where they are stronger) and corona, traversing several mega-meters and cause (quasi)steady heating in the mossy plage regions. Dissipation of the current sheets leads to strong heating of the plasma in the chromosphere and the corona that explains the cause of the observed correlation. These types of small-scale current sheets from braiding are more ubiquitous and very different from large-scale current sheets that form when newly emerging flux interacts with pre-existing fields and that have been implicated in local heating\cite{2022A&A...661A..59D}. The current sheets in our simulation occur on very small-scales of tens of km that are at (and beyond) the limit of current observational capabilities\cite{2021ApJ...921...39A}. High-resolution magnetic field measurements at multiple heights are needed to resolve some of these sheets. Future observations, e.g. with 4-m DKIST\cite{2020SoPh..295..172R}, may aid in this direction.  


Furthermore, the nature of heating observed in the moss appears to be incompatible with current models of wave-based heating. Predicted heating rates from Alfv\'en wave turbulence models\cite{2014ApJ...787...87V,2017ApJ...849...46V} are highly time varying because of the presence of high frequency waves. This causes a lack of correlation between chromospheric and upper TR (lower coronal) emission, in contrast to what we observe in this study. Future observations of longer duration with Solar Orbiter's Extreme Ultraviolet Imager\cite{2020A&A...642A...8R} instrument and the upcoming NASA MUlti-slit Solar Explorer (MUSE) mission\cite{2020ApJ...888....3D} will provide more detailed imaging and spectroscopy of a much wider variety of regions, and help determine how widespread this heating mechanism is in the solar atmosphere.


\bibliography{sample}

\section*{Materials and Methods}
\label{sec:methods}

\subsection*{Observations}
The Hi-C 2.1 sounding rocket mission was launched on 29 May 2018 at 18:54~UT from the White Sands Missile Range in New Mexico, USA. The target was AR 12712 and the instrument successfully acquired high spatial resolution (0.4"), 2k$\times$2k images with a 2s exposure at a cadence of 4.4s. The Solar Pointing and Aerobee Control System (SPARCS) maintained a constant target for the entire duration of the flight lasting 335s. One of the major goals of this mission was to study the mass and energy coupling between the chromosphere and the corona in ARs. Consequently, the observations were recorded in the upper transition/lower coronal 172~\AA\ passband (\ion{Fe}{ix}/\ion{Fe}{x}) that has a peak temperature sensitivity of about 1~MK. The passband is similar to AIA 171~\AA\ channel, except that Hi-C 2.1 recorded observations at roughly three times the spatial and temporal resolution of AIA thereby allowing an unprecedented view of the million degree corona. The details of the instrument and its characterization can be found here\cite{2019SoPh..294..174R}.    

The Hi-C 2.1 sounding rocket flight was also coordinated with IRIS that co-observed the AR target with multiple, co-temporal, high-cadence rasters centered around the moss brightenings (indicated by the yellow dashed lines in Fig.~\ref{fig:context}~b). It ran in very large, sparse 8-step raster mode (OBS-3600104031) with a step size of 1" perpendicular to the slit and a raster cadence of 25~s covering a FOV of 175"$\times$7" in the solar $Y$ and $X$ directions. 256 rasters were obtained over a duration of 1hr and 50 min, including the Hi-C flight duration of 5 mins (equivalent to 14 rasters). In addition, IRIS also recorded 
several very large, dense 400-step rasters (OBS-3600010078) covering a FOV of 141"$\times$174" co-spatially with Hi-C, but for roughly 1hr and 51~min (raster cadence) before and after the sounding rocket flight. The step size in this case was 0.35" perpendicular to the slit, while the step cadence was 16.7s. The FOV of one such context raster is indicated in Fig.~\ref{fig:context}~(b) and the purpose of these rasters was to provide contextual information in and around the AR. Furthermore, IRIS recorded slit-jaw images (SJIs) in \ion{C}{ii}~1330~\AA, \Si{}~1400~\AA, \ion{Mg}{ii}~2796~\AA, and Mg continuum 2832~\AA\ channels sampling plasma from 6,000 to $\sim$100,000 K. The cadence and the spatial sampling of these images are 13s and 0.33", respectively.

Finally, we have also used observations in the SDO\citeMethods{2012SoPh..275....3P}/AIA\citeMethods{2012SoPh..275...17L} 94~\AA\ (\ion{Fe}{XVIII},$\sim$6~MK) and 335~\AA\ (\ion{Fe}{XVI},$\sim$2.5~MK) passbands solely for the purpose of visualizing the presence of hot AR loops above the moss brightenings seen in \HiC{}. This relationship is shown in the form an RGB composite image in Fig.~\ref{fig:context}~(b).

\subsection*{Data processing}

The above sets of observations were precisely co-aligned using SolarSoftWare (SSW\citeMethods{1998SoPh..182..497F}) IDL routines including compensating for the roll angle (1.7 degrees) between the IRIS and Hi-C observations. In particular, the IRIS SJIs and Hi-C datasets were co-aligned through first, a visual comparison and cross-correlation of \HiC{} and AIA 171~\AA\ channels, followed by internally co-aligning the AIA 1600~\AA, 171~\AA, 94~\AA, and 335~\AA\ channels (this includes the 3hr cadence limb fitting), and finally by co-aligning the AIA 1600~\AA\ and IRIS 1400~\AA\ SJIs. The above steps resulted in a precise co-alignment between IRIS SJIs and the Hi-C 2.1 data, where the former was expanded\citeMethods{park1983image}, rotated and spatio-temporally aligned to the latter. Extensive visualization with CRISPEX\citeMethods{2012ApJ...750...22V}, an IDL widget-based tool, showed good correspondence among the coordinated datasets.

For the spatio-temporal correlation analysis presented in the paper, synthetic sparse, 8-step \HiC{} rasters were generated that were spatially and temporally co-aligned with each of the IRIS rasters. This was done to ensure consistency between the IRIS and Hi-C observations, since the spectrograph slit takes time to ``build" the FOV whereas Hi-C observations over the whole FOV are instantaneous. It is to be noted that the derived parameters (e.g. intensity of \kcore{}, \Si{}, $T$, N$_{\mathrm{e}}$) from the IRIS rasters at original resolution were resampled to Hi-C 2.1 pixel scale (0.129") before generating the synthetic Hi-C rasters, which was done to not compromise the high spatial resolution of Hi-C. A standard single Gaussian fitting algorithm was applied to derive the \Si{} peak intensity, while a double Gaussian approach (including one absorption and emission Gaussian like ref.\citeMethods{2015ApJ...811..127S}) was employed to fit the \Mgk{} spectra. This way single peaked profiles were easily fitted with the emission Gaussian whereas the remaining spectra with depressed cores were fitted with a superposition of a broad emission and narrower absorption Gaussian.

\subsection*{$\mathbf{IRIS^{2}}$ inversions}

The inversions presented in this article were made using the \irissq\ inversion framework. The core of this framework is a database created from the inversion of multi-line {\it representative profiles} (RPs) and their corresponding {\it representative model atmospheres} (RMAs). 

This approached was first introduced by Sainz Dalda et al.\cite{2019ApJ...875L..18S}. There, the authors calculated the RPs of the \ion{Mg}{ii}~h\&k lines (including the \ion{Mg}{ii}~UV lines located between the former ones) for a selection of IRIS data sets. The RPs are the averaged profile of those profiles belonging to a cluster. A cluster is calculated by using the $k-means$ technique\citeMethods{Steinhaus57,macqueen1967}. The profiles within a cluster share a similar shape, i.e. the distribution of the intensity (spectral radiance) with respect to the wavelength is similar. Therefore, the physical conditions of the atmosphere where the radiation comes from are similar in the profiles within a cluster, and it is encoded in the RP of the cluster. 
The database was created considering a large variety of RPs belonging to different solar features with different observational setups, e.g., exposure time, location on the solar disk, raster steps, and so on.
These RPs were inverted by the state-of-the-art STockholm Inversion Code (STiC\citeMethods{2019A&A...623A..74D}), which considers non-LTE radiative transfer and partial frequency distribution (PRD) of the scattered photons. This code is capable of inverting simultaneously several spectral lines from different atomic species, both in non-LTE and LTE. STiC solves the radiative transfer problem to synthesize the spectral lines in the RP, and in an iterative process that slightly changes the physical parameters of a initial guess model, looks for the best possible fit between the synthetic profile and the observed RP. This process is repeated until convergence under a given limit or computational constraint.  As a result of the inversion, STiC provides a synthetic RP that best fits the observed RP, and the RMA that reproduces that synthetic RP. An RMA consists of thermodynamic parameters (e.g., $T$, N$_{\mathrm{e}}$, LOS velocity, microturbulence) as a function of optical depth ($\tau$).
For a given observed \ion{Mg}{ii}~h\&k profile in an IRIS observation, \irissq\ looks for the closest synthetic RP in their database that fits that observed profile, and associates the RMA to the location in the data set where the observed profile was recorded. This method speeds the inversion of a data set up by a factor 10$^{5}$--10$^{6}$. 

In this paper, we use an extended version of the \irissq\ method. Sainz Dalda et al.~2022~(in prep) use the same method and calculate the RPs for multi-line profiles containing the \ion{C}{ii} lines, the \ion{Mg}{ii}~h\&k lines, the three lines of the \ion{Mg}{ii}~UV triplet, and 6 photospheric lines located around the \ion{Mg}{ii}~h\&k spectral range. Thus, a multi-line RP is simultaneously inverted by STiC providing a RMA, for which the thermodynamic parameters are constrained from the top of the chromosphere to the bottom of the photosphere (see Sainz Dalda et al.~2022~in prep for more details).
Note that because we are considering simultaneously several lines that are sensitive to variations of the thermodynamics in the same region of the solar atmosphere, the $T$ and the non-thermal contribution to the profile (microturbulence) are better constrained than in the case that only considers one spectral line, which is the case for both the chromosphere and the low photosphere. Thus, this set of RPs and RMAs are likely the best constrained, comprehensive database of thermodynamics models of the low solar atmosphere. 

For this paper, \irissq\ looked for the closest RP in the database considering only the lines available in our data sets, which are the \ion{C}{ii} lines, the \ion{Mg}{ii}~h\&k lines, the three \ion{Mg}{ii}~UV triplet lines, and the photospheric \ion{Ni}{i} at 2815.18 \AA line.
Fig.~\ref{fig:supp_inv_ex1} shows the observed lines in the chromosphere (red-dotted line) and their fits (black).
Thus, although the RMA provided by \irissq\ contains information that was obtained from inverting 6 photospheric lines, the information presented in this paper comes from fitting only 1 photospheric line. Therefore, the values in the region log$\tau$=[-2,0] must be considered with caution. 
We are however very confident with the values obtained in the chromosphere, especially between log$\tau$=[$-$6,~$-$2], since that information comes from the unique simultaneous inversion of 6 chromospheric lines, which provide model atmospheres with excellent constraint in these heights, and the photospheric \ion{Ni}{i} at 2815.18 \AA line.

\subsection*{Numerical Simulation}
\begin{figure}[ht!]
\centering
\includegraphics[scale=0.8]{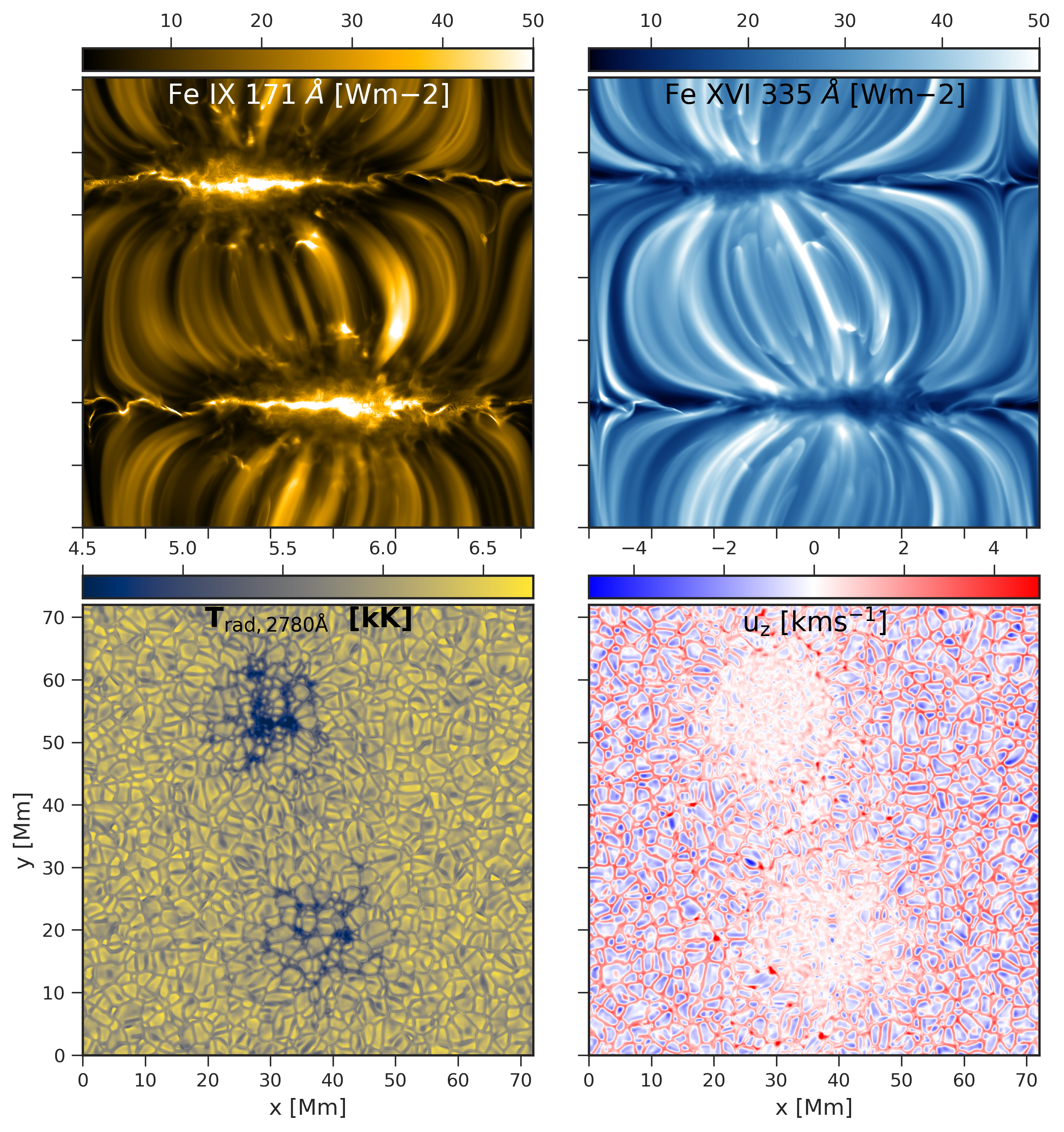}
\caption{Coronal and photospheric scenery of the 3D MHD plage simulation. \textit{Top left:} Synthetic \ion{Fe}{ix}~171~\AA\ emission showing bright, reticulated moss patches. \textit{Top right:} Synthetic \ion{Fe}{xvi}~335~\AA\ emisison showing hot (2.5~MK) loops overlying the moss. \textit{Lower left:} Synthetic photospheric radiation temperature at 2780~\AA. \textit{Lower right:} Vertical velocity (u$_{\mathrm{z}}$) map at $z$=0~Mm.}
\label{fig:sim_hot_loops}
\end{figure}

The 3D simulation of the plage-like region was performed in a computational box with dimensions $72\times72\times61$~Mm spanning from 8.5~Mm below the photosphere to 52.5~Mm above it. The resolution of this box is moderate: 100~km in the horizontal directions and $[20,100]$~km in the vertical direction depending on location in the atmosphere; in the photosphere, chromosphere and transition region the vertical resolution is of order 20~km. The model calculations are run with the Bifrost radiative-MHD code\citeMethods{2011A&A...531A.154G}, which includes optically thick radiative transfer in the photosphere and lower chromosphere, effectively thin radiative losses in the mid- and upper chromosphere, and optically thin radiative losses in the corona. Spitzer thermal conduction along the magnetic field is handled by a multi-grid method (or alternately by the hyperbolic method developed by ref.\citeMethods{2018ApJ...859..161R}) to maintain time steps at a reasonable level. To initialize the model, two circular patches of strong but opposite polarity magnetic flux (at 8.5~Mm below the photosphere) were used to generate a potential field covering the entire computational box. This field geometry was inserted into a relaxed convective atmosphere after which the field and atmosphere were allowed to relax together. At the time of the snapshot the photosphere in the strong field regions has become quite cold (and darker than the surrounding atmosphere) while the chromosphere and corona are hot and bright (Fig.~\ref{fig:sim_analysis} and Supp. Fig.~\ref{fig:sim_hot_loops}).


In order to compute the \ion{Mg}{ii} lines we have used the RH 1.5D code\citeMethods{2001ApJ...557..389U,2015A&A...574A...3P}. This code includes detailed radiative physics including polarized transfer and PRD, but at the cost of only being used for 1.5D (column-by-column) computations, and thus ignores the 3D effects of lateral transport of radiation. 3D effects can play an important role in modeling the \ion{Mg}{ii} lines\cite{2013ApJ...772...90L}, especially in the line cores. However, in regions where the temperature and the density of the plasma is enhanced (such as in the mossy plage), 3D effects can be minimal due to lower scattering of photons. Moreover, the lower time complexity of the 1.5D approach far outweighs its limitations. The calculations are carried out in non-LTE assuming PRD.


The emission of optically thin transition region and coronal lines, including the \ion{Fe}{ix}~17.1~nm and \ion{Fe}{xvi}~33.5~nm lines are calculated using contribution functions $G(T,n_{\rm e})$ from the CHIANTI\citeMethods{2015A&A...582A..56D} database package. These contribution functions are integrated along the line of sight using the local values of the electron density $n_{\rm e}$, the temperature $T$, and the velocity along the line of sight u$_z$ at several frequencies covering the emitting ions line profile. These calculations are done on GPUs using a Cuda code developed by Juan Martinez-Sykora.   

\bibliographystyleMethods{naturemag-doi}
\bibliographyMethods{supplementary}

\section*{Acknowledgments}

S.B., V.H., A.S.D., and B.D.P. gratefully acknowledge support from NASA contract NNG09FA40C (IRIS). B.D.P. was also supported by grant NNM16AA10P (Hi-C). We acknowledge the High-resolution Coronal Imager (Hi-C 2.1) instrument team for making the second re-flight data available under NASA proposal 17-HTIDS17\_2-003. MSFC/NASA led the mission with partners including the Smithsonian Astrophysical Observatory, the University of Central Lancashire, and the Lockheed Martin Solar and Astrophysics Laboratory. Hi-C 2.1 was launched out of the White Sands Missile Range on 2018 May 29. IRIS is a NASA small explorer mission developed and operated by LMSAL with mission operations executed at NASA Ames Research Center and major contributions to downlink communications funded by ESA and the Norwegian Space Centre. The 3D visualization of the numerical simulation was carried out using VAPOR, a product of the Computational Information Systems Laboratory at the National Center for Atmospheric Research.

\section*{Author contributions statement}
S.B. was responsible for most of the analysis and writing of the manuscript. B.D.P. designed the study, performed initial analysis, and final editing. V.H. performed the numerical simulations, calculation of synthetic observables, and assisted with analysis of the simulations. A.S.D. performed the inversions. S.S. and A.W. designed, built, and operated the Hi-C sounding rocket experiment. All authors reviewed the manuscript.

\section*{Competing Interests}
The authors declare no competing interests.


\section*{Supplementary Section}
\label{sec:supplements}

\end{document}